\documentclass{article}
\usepackage{PRIMEarxiv}
\usepackage[utf8]{inputenc}
\usepackage[T1]{fontenc}
\usepackage{amsmath}
\usepackage{amsfonts}
\usepackage{graphicx}
\graphicspath{{media/}}
\usepackage{booktabs}
\usepackage{array}
\usepackage{longtable}
\usepackage{tabularx}
\usepackage{tikz}
\usetikzlibrary{arrows.meta}
\usetikzlibrary{positioning}
\usepackage{subcaption}
\usepackage{hyperref}
\usepackage{url}
\usepackage{microtype}
\usepackage{nicefrac}
\usepackage{todonotes}
\usepackage{ifthen}
\usepackage[misc]{ifsym}
\usepackage{lipsum}
\usepackage{listings}
\usepackage{xcolor}
\usepackage{fancyhdr}
\newif\ifcomments
\commentstrue      
\ifcomments
  \newcommand{\ivan}[1]{\textcolor{blue}{\textbf{[Iván:} #1\textbf{]}}}
  \newcommand{\armen}[1]{\textcolor{teal}{\textbf{[Armen:} #1\textbf{]}}}
  \newcommand{\aaron}[1]{\textcolor{magenta}{\textbf{[Aaron:} #1\textbf{]}}}
  \newcommand{\jordi}[1]{\textcolor{red}{\textbf{[Jordi:} #1\textbf{]}}}
\else
  \newcommand{\ivan}[1]{}
  \newcommand{\armen}[1]{}
  \newcommand{\aaron}[1]{}
  \newcommand{\jordi}[1]{}
\fi

\title{Low-code and no-code with BESSER to create and deploy smart web applications}

\author{
Iván Alfonso \\
Luxembourg Institute of Science and Technology, Luxembourg \\
\texttt{ivan.alfonso@list.lu}
\and
Armen Sulejmani \\
Luxembourg Institute of Science and Technology, Luxembourg \\
\texttt{armen.sulejmani@list.lu}
\and
Aaron Conrardy \\
Luxembourg Institute of Science and Technology, Luxembourg \\
University of Luxembourg, Luxembourg \\
\texttt{aaron.conrardy@list.lu}
\and
Jordi Cabot \\
Luxembourg Institute of Science and Technology, Luxembourg \\
University of Luxembourg, Luxembourg \\
\texttt{jordi.cabot@list.lu}
}

\begin{document}
\newpage
\thispagestyle{empty}
\vspace*{\fill}
\begin{center}
\Large
\textbf{This is a preprint.}\\[1em]
The peer-reviewed paper is published in the proceedings of ICWE 2026 as part of the demo track.\\[1em]
Please reference the published version. \\[1em]

DOI: \url{https://doi.org/10.1007/978-3-032-29372-5_32}
\end{center}
\vspace*{\fill}
\newpage
\maketitle

\begin{abstract}
The increasing demand for web applications containing AI-agents, seen as smart web applications, has prompted the need for new techniques to facilitate their creation. Low-code has risen as an approach that reduces the amount of handwritten code by focusing on the abstraction of components in the form of models combined with automated generators to produce applications. Existing low-code platforms are commercial, leading to drawbacks such as the risk of vendor lock-in, limited extensibility, and more. We present the open-source BESSER low-code framework, which allows users to design, generate and deploy their application via a freely accessible web-based editor, while guaranteeing transparency and extensibility.
\end{abstract}


\section{Introduction}


The increasing demand for better software in recent years has created a need for advanced strategies to improve developer productivity. Low-code development, the latest reincarnation 
of model-driven engineering approaches, addresses this demand but introduces 
challenges related to the growing complexity of modern software requirements, such as 
AI-powered agents.
Low-code development platforms (LCDPs) aim to accelerate software development by relying on high-level models and automated code generation, 
reducing manual programming effort.

However, while these platforms are effective for building conventional applications and even improved their support for smart components, the most popular and complete ones are usually the commercial ones \cite{interoperability}. 
This leads to drawbacks such as limited support for interoperability, vendor lock-in, limited choice for technological stack for external components or deployment, limited to no extensibility, and limited to no access to generated code.

On the other hand, applying model-driven approaches supported by graphical notations to the development of web and mobile user interfaces (UIs) is nothing new. Languages such as IFML \cite{ifml}, OOWS \cite{Fons2008}, or UWE \cite{uwe} are often seen as foundational for the development of traditional UIs. Yet, these do not cover the definition of smart components, and provide limited or fragmented tool support for the complete creation and deployment pipeline.


In this paper, we present a demo of the open-source low-code platform BESSER \cite{alfonso2024building}, which aims to address the challenges described above. Specifically, BESSER offers a freely available web-based modeling editor\footnote{\url{https://github.com/BESSER-PEARL}} for design and deployment of smart web applications following a low-code and no-code approach.

\section{Demonstration}

\begin{figure}[t]
    \centering
    \includegraphics[width=1\linewidth]{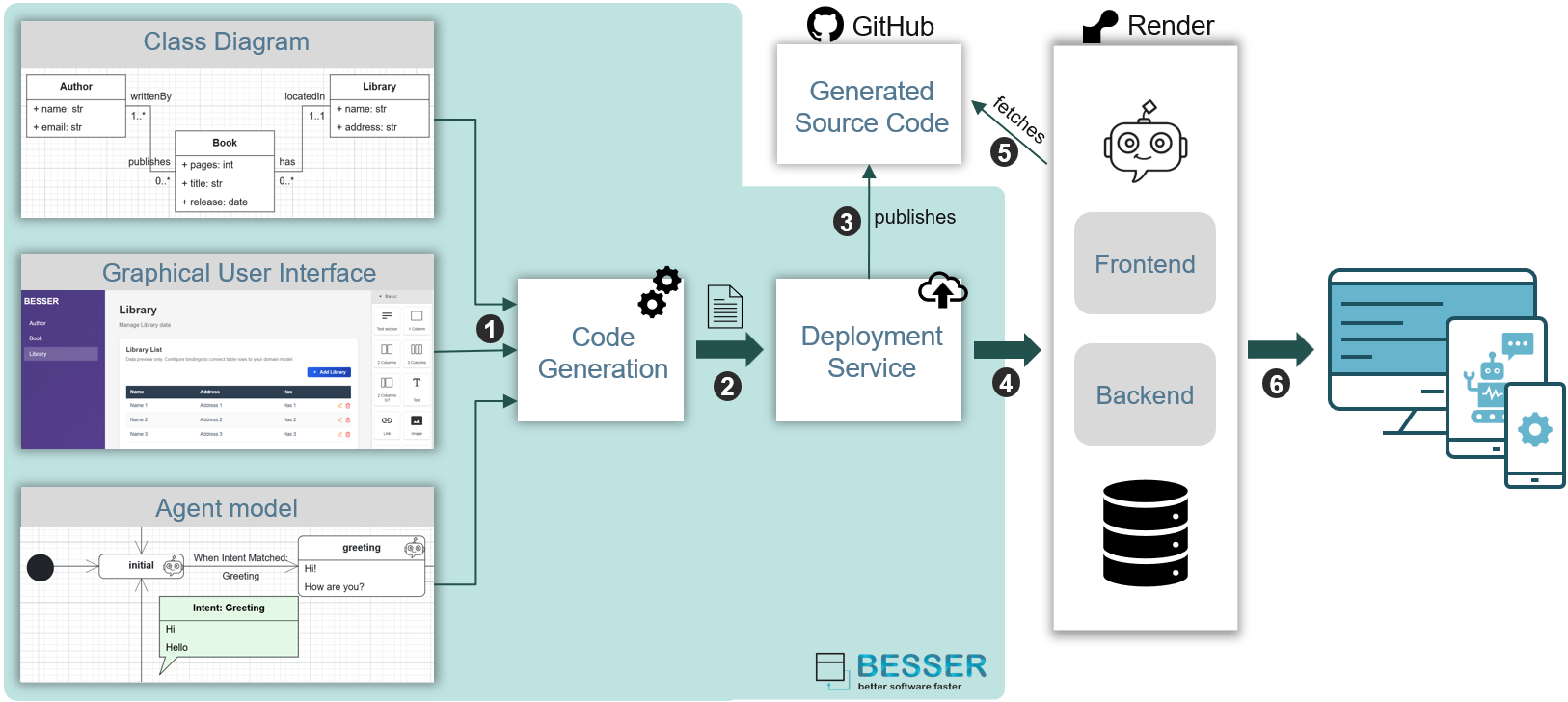}
    \caption{BESSER pipeline for smart web applications}
    \label{fig:besser}
\end{figure}

Figure \ref{fig:besser} presents the part of BESSER's pipeline for developing smart web applications used in the demonstration. 
The demonstration contains the example of a web application of a library, with a conversational agent responding to frequently asked questions. During the demo, this scenario is specified through three models: a class diagram, a graphical user interface (GUI), and an agent model. These models are provided as a template on the BESSER platform (\url{https://editor.besser-pearl.org}), allowing attendees to load them and then generate and deploy the application. The only additional requirement to complete the demo is a free account on GitHub\footnote{https://github.com/} and Render\footnote{https://render.com/}, the former being used for hosting the models and generated code, and the latter for running frontend and backend servers. A video demonstration of the presented scenario is available\footnote{\url{https://besser.readthedocs.io/en/latest/examples/smart_web.html}}


\begin{figure}[t]
    \centering
    \includegraphics[width=\linewidth]{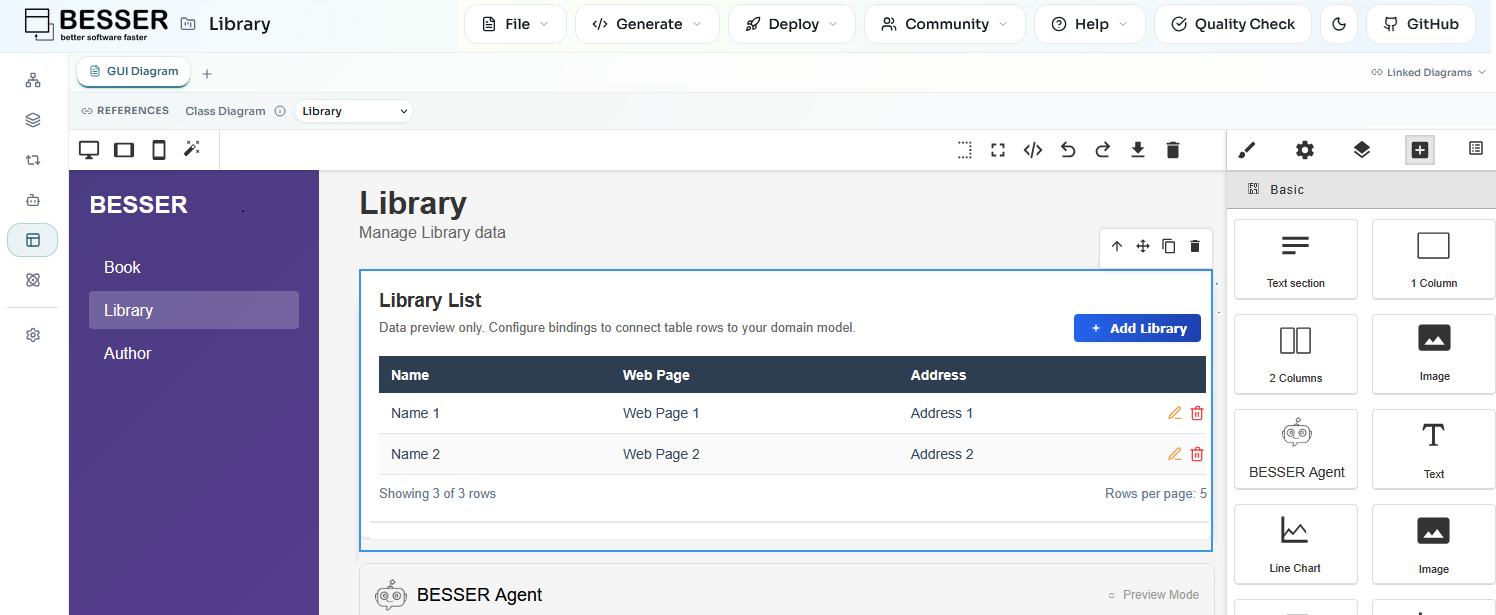}
    \caption{GUI editor in BESSER}
    \label{fig:gui_part}
\end{figure}
\section{Modeling, generating and deploying smart web applications with BESSER}

BESSER provides the B-UML (BESSER-UML) language, which enables the modeling of smart web applications via three complementary but still interrelated perspectives: \textit{structural}, \textit{agent}, and \textit{GUI} perspectives.

The \textbf{structural} perspective specifies the static aspects of the web application through class diagrams, including classes, attributes, methods, relationships, and other elements, reusing core UML \cite{uml2017omg} concepts for this perspective while extending and simplifying selected aspects of the specification for smart web applications. For example, meta-attributes such as \texttt{description}, \texttt{URI}, and \texttt{icon} are added to classes to capture platform-specific information.
The \textbf{agent} perspective enables explicit modeling of autonomous agents using an extended state-machine language. 
Agents are defined in terms of states, transitions, intents, and actions executed in each state. These actions may include textual responses (e.g., for chat interactions), calls to Python methods (e.g., to invoke external APIs), or operations driven by Large Language Model (LLM) outputs, combining the controlled output from state machines with the openness of LLMs.

From the \textbf{GUI} perspective, the editor provides a no-code interface that allows users to drag and drop UI components onto a blank page (Figure \ref{fig:gui_part}), which are immediately rendered. The designed page is internally parsed into the GUI part of B-UML, inspired by IFML \cite{ifml} and extended with constructs for rich view components (e.g., charts and chat interfaces) and styling properties (e.g., colors and layout). The GUI can link to the structural and agent models, consolidating all models into a smart web application.






The generation pipeline orchestrates specialized sub-generators, each responsible for a distinct technology layer, producing a deployment-ready project from the modeled smart web application.

The \textbf{backend generator} transforms the structural model into three interconnected Python modules: \textit{Pydantic classes} for data validation, \textit{SQLAlchemy models} for database schema and ORM, and a \textit{FastAPI REST API} exposing comprehensive CRUD endpoints for all entities and relationships. 
The \textbf{frontend generator} interprets the GUI model to produce a React-based application with TypeScript. View components are mapped to React components including data tables, interactive forms, method invocation buttons, and chart visualizations. 
When an \textbf{agent model} is provided, \textbf{BESSER}'s \textbf{Agentic Framework (BAF) generator} generates the agent backend as executable Python scripts. The generated backend exposes a WebSocket endpoint, and the frontend embeds a chat widget that connects to this service to exchange messages and display responses.


BESSER streamlines cloud deployment by integrating with GitHub and Render for one-click deployment. Users authenticate via GitHub OAuth and trigger deployment from the editor. The backend generates the application, creates a GitHub repository, and pushes the codebase with a \texttt{render.yaml} configuration defining free-tier services: FastAPI backend, React frontend, and optional agent service. Render automatically detects this configuration, provisions services, installs dependencies, and deploys all components. The platform provides live URLs with SSL certificates and manages cold-start behavior. This workflow enables users to go from model edits to a publicly accessible web application typically within a few minutes with no manual infrastructure configuration.

\section{Conclusion}
In this paper, we showcase how BESSER enables the design, generation and deployment of a smart web application following a low-code / no-code approach through a web-based editor. We described the architecture and components relevant to the demonstration, chosen to enable extensibility. 
The generated code can be deployed as is or act as a prototype that can be manually extended. For the future, we plan to explore how perspectives can be integrated within the creation of smart web applications, such as the integration of personalization via user profiles or recommendation components using neural networks.

\section*{Acknowledgments}
This work is supported by the Luxembourg National Research Fund (FNR) PEARL program, grant agreement 16544475.

\bibliographystyle{unsrt}
\bibliography{bib}

@inproceedings{alfonso2024building,
  title={Building {BESSER}: an open-source low-code platform},
  author={Alfonso, Iv{\'a}n and Conrardy, Aaron and Sulejmani, Armen and Nirumand, Atefeh and Ul Haq, Fitash and Gomez-Vazquez, Marcos and Sottet, Jean-S{\'e}bastien and Cabot, Jordi},
  booktitle={International Conference on Business Process Modeling, Development and Support},
  pages={203--212},
  year={2024},
  organization={Springer}
}

@InProceedings{interoperability,
author="Alfonso, Iv{\'a}n
and Conrardy, Aaron
and Cabot, Jordi",
title="Towards the Interoperability of Low-Code Platforms",
booktitle="Intelligent Information Systems",
year="2025",
publisher="Springer Nature Switzerland",
address="Cham",
pages="3--11"
}

@article{uml2017omg,
  author = {{Object Management Group}},
  title = {{OMG} Unified Modeling Language ({OMG UML}), Version 2.5.1},
  year = {2017},
  howpublished = {\url{https://www.omg.org/spec/UML/2.5.1}},
  note = {{Object Management Group (OMG) Standard}}
}

@book{ifml,
  title={Interaction flow modeling language: Model-driven UI engineering of web and mobile apps with IFML},
  author={Brambilla, Marco and Fraternali, Piero},
  year={2014},
  publisher={Morgan Kaufmann}
}

@InProceedings{uwe,
  author    = "Hennicker, Rolf and Koch, Nora",
  editor    = "Evans, Andy and Kent, Stuart and Selic, Bran",
  title     = "A UML-Based Methodology for Hypermedia Design",
  booktitle = "{\guillemotleft}UML{\guillemotright} 2000 --- The Unified Modeling Language",
  year      = "2000",
  publisher = "Springer Berlin Heidelberg",
  address   = "Berlin, Heidelberg",
  pages     = "410--424",
}

@Inbook{Fons2008,
author="Fons, Joan
and Pelechano, Vicente
and Pastor, Oscar
and Valderas, Pedro
and Torres, Victoria",
title="Applying the Oows Model-Driven Approach for Developing Web Applications. The Internet Movie Database Case Study",
bookTitle="Web Engineering: Modelling and Implementing Web Applications",
year="2008",
publisher="Springer London",
pages="65--108",
}

\end{document}